\documentclass[preprint,12pt]{elsarticle}

\usepackage{amsmath}
\usepackage{amsfonts}
\usepackage{amsthm}
\usepackage{amssymb}
\usepackage{graphicx}

\newtheorem*{Assumption}{Assumption}

 \makeatletter
    
    \@addtoreset{equation}{section}
  \makeatother

\begin{document}

\begin{frontmatter}

\title{Turing instability in Reaction-Diffusion models on complex networks}

\author[ide]{Yusuke Ide}
\address[ide]{Faculty of Engineering, Kanagawa University, 3-27-1, Rokkakubashi, Kanagawa-ku, Yokohama, Kanagawa, 221-8686, Japan}
\author[izuhara]{Hirofumi Izuhara}
\address[izuhara]{Faculty of Engineering, University of Miyazaki, 1-1, Gakuen Kibanadai Nishi, Miyazaki, 889-2192, Japan}
\author[machida]{Takuya Machida}
\address[machida]{Japan Society for the Promotion of Science, Japan}

\begin{abstract}
In this paper, the Turing instability in reaction-diffusion models defined on complex networks is studied. 
Here, we focus on three types of models which generate complex networks, i.e. the Erd\H{o}s-R\'enyi, the Watts-Strogatz, and the threshold network models. 
From analysis of the Laplacian matrices of graphs generated by these models, we numerically reveal that stable and unstable regions of a homogeneous steady state on the parameter space of two diffusion coefficients completely differ, depending on the network architecture. 
In addition, we theoretically discuss the stable and unstable regions in the cases of regular enhanced ring lattices which include regular circles, and networks generated by the threshold network model when the number of vertices is large enough. 
\end{abstract}

\begin{keyword}

Turing instability; Reaction-diffusion models on networks; Complex network; Pattern formation
\end{keyword}

\end{frontmatter}

\section{Introduction}
\label{Introduction}
We can observe various types of pattern phenomena in nature. 
In order to understand the pattern formation mechanisms of such phenomena, mathematical models have been proposed and analyzed from the viewpoint of both numerical and theoretical studies. 
Among these models, reaction-diffusion systems have attracted many researchers \cite{M}. 
Though the systems which describe a local interaction and a long-range dispersal between chemical substances or biological species are rather simple, Turing stated that spatially inhomogeneous structures can be formed in a self-organized way under certain conditions \cite{T}. 
Since then, a lot of studies on the reaction-diffusion systems have been reported. Turing considered the following system of partial differential equations: 
\begin{equation}
\label{RD}
\begin{aligned}
\frac{\partial u}{\partial t}&=d_u\Delta u +f(u ,v ),\\
\frac{\partial v}{\partial t}&=d_v\Delta v +g(u ,v ),
\end{aligned}
\end{equation}
where $u=u(t,x)$ and $v=v(t,x)$ indicate concentrations of chemical substances or population densities of biological species at time $t$ at position $x$, $d_u$ and $d_v$ mean respectively diffusion coefficients of $u$ and $v$, and the functions $f(u,v)$ and $g(u,v)$ express a local interaction between $u$ and $v$. 
In addition, Turing gave the following assumption on the reaction system of ordinary differential equations without diffusion terms 
\begin{equation}
\label{ODE}
\begin{aligned}
\frac{d u}{d t}&=f(u,v), \\
\frac{d v}{d t}&=g(u,v). 
\end{aligned}
\end{equation}
\begin{Assumption}
The system \eqref{ODE} possesses an equilibrium point $(u,v)=(\overline{u},\overline{v})$ and it is asymptotically stable. 
\end{Assumption}

In this framework, Turing derived a paradox that the equilibrium solution $(u,v)=(\overline{u},\overline{v})$ in \eqref{RD} with suitable boundary conditions can be destabilized in spite of adding the diffusion terms which possess a smoothing effect of spatial heterogeneity even though the equilibrium point $(u,v)=(\overline{u},\overline{v})$ is stable in the sense of \eqref{ODE}. This is well known as the diffusion-induced instability or the Turing instability. As a consequence of the Turing instability, \eqref{RD} exhibits spatially inhomogeneous structures called Turing patterns. Therefore, the Turing instability is regarded as important for the onset of pattern formation on the reaction-diffusion systems. 

However, since \eqref{RD} is a system describing a local interaction and a dispersal between $u$ and $v$ on continuous media, it can not represent an interaction on a spatially discrete environment, such as dynamics of metapopulation, cellular networks of biological morphogenesis and networks of diffusively coupled chemical reactors. Therefore, in order to treat such situations, studies on reaction-diffusion models defined on networks have proceeded \cite{OS,HLM,MH,AF,NM,W, FA, HNM, KHG, ACPSF, ABCFP, HNM2, APF}. 
So far, the Turing instability arising in reaction-diffusion models defined on  networks with a small number of vertices has been investigated \cite{OS, HLM, MH,AF}. Recently, studies on Turing patterns formed on complex networks with a large number of vertices have proceeded \cite{NM,W,FA, HNM, KHG, ACPSF, ABCFP, HNM2, APF}. In these papers, the differences between the classical Turing patterns on continuous media and network organized Turing patterns, and properties on Turing patterns on complex networks have been shown by numerical simulations, a theory of mean field approximation, and analytical techniques. The authors in \cite{NM,W,FA, HNM2} also discussed the Turing instability on networks with a large number of vertices in a somewhat general framework, which the Turing patterns are based on. (For example, see the Methods section in \cite{NM}.) However, they do not mention the influence of network topology on the Turing instability. In other words, it is not clear whether the Turing instability on a complex network occurs  or not when the network topology changes. In this study, considering reaction-diffsuion models on complex networks with a large number of vertices, we investigate a relation between the Turing instability and network topology with a large number of vertices in detail by using the linear stability analysis. 
In particular, our interest is how network topology influences the Turing instability. Therefore, we focus on three types of models which generate typical networks, i.e. the Erd\H{o}s-R\'enyi, the Watts-Strogatz, and the threshold network models. We emphasize that the Turing instability is an important concept as the onset of self-organized pattern formation. However, thorough studies focusing on the Turing instability in reaction-diffusion models on complex networks with many vertices are very few. 
Potential applications of reaction-diffusion models on networks were introduced in \cite{NM, FA,HNM, PV}. 
If we regard vertices as cells, the model describes cellular networks of early biological morphogenesis. When vertices are regarded as patchy habitats, it is a model describing population dynamics with an interaction on a patchy environment (ecological metapopulation). In order to understand self-organized phenomena with the network architecture, a theoretical approach using models defined on networks could be needed. As a first step, we discuss the Turing instability on complex networks in this paper.

This paper is organized as follows: in the next section, we formulate reaction-diffusion models defined on complex networks, which we discuss in this paper. Sections \ref{Networks generated by Erdos-Renyi model}, \ref{Networks generated by Watts-Strogatz model}, and \ref{Networks generated by Threshold model} are devoted to computer-aided analysis of the Turing instability in reaction-diffusion models on networks generated by the Erd\H{o}s-R\'enyi, the Watts-Strogatz, and the threshold network models, respectively. We reveal that these analyses derive different results on the Turing instability of a homogeneous steady state, depending on network topology. In section \ref{Theoretical results}, we give theoretical results on the instability when the number of vertices is large enough. We complete this paper in section \ref{Concluding remarks}, where concluding remarks and future works are listed.

\section{Formulation of a reaction-diffusion model on a graph}
\label{Formulation of Reaction-Diffusion model on networks}

As an analogy of the reaction-diffusion model on continuous media \eqref{RD}, we formulate a model on a graph with $N$ vertices. 
When there is no connection between any vertices, the dynamics on each vertex is described by a local interaction only as follows:  
\begin{equation}
\label{dynamics vertex}
\begin{aligned}
\frac{du_i}{dt}&= f(u_i ,v_i ),\\
\frac{dv_i}{dt}&= g(u_i ,v_i ), 
\end{aligned}
\qquad (i=1,2,\cdots, N),
\end{equation}
where a pair $(u_i,v_i)=(u_i(t),v_i(t))$ denotes some quantities on the $i$th vertex, such as population densities of biological species or concentrations of chemical substances. 
Note that the interaction functions $f$ and $g$ are the same on each vertex. 
As well as the Turing instability on continuous media, we assume the existence of an asymptotically stable equilibrium point $(u_i,v_i)=(\overline{u},\overline{v})$ $(i=1,\cdots,N)$ for \eqref{dynamics vertex}. 
The asymptotic stability of the equilibrium point means that 
\begin{equation}
\label{parameter stable}
a+d<0 \text{ and } a d-b c>0 
\end{equation}
from information of the linearized system of \eqref{dynamics vertex} around $(u_i,v_i)=(\overline{u},\overline{v})$, where $a=f_{u_i}(\overline{u},\overline{v})$, $b=f_{v_i}(\overline{u},\overline{v})$, $c=g_{u_i}(\overline{u},\overline{v})$ and $d=g_{v_i}(\overline{u},\overline{v})$. Next, when connection between vertices is taken into account, we give an assumption on a flux of the quantities between vertices. If the $i$th and the $j$th vertices are connected by an edge, then we suppose that a flux exists between these vertices. On the other hand, if two vertices are not connected, there is no flux between them. We assume that the flux is given by Fick's law of diffusion, which means that the flux is proportional to the difference of quantities on the two vertices. Therefore, the dynamics of $u_i$ and $v_i$ on the $i$th vertex is described as 
\begin{equation}
\label{NRD}
\begin{aligned}
\frac{du_i}{dt}&= d_u\sum_{j=1}^{N} A_{ij} (u_j-u_i)+f(u_i,v_i),\\
\frac{dv_i}{dt}&= d_v\sum_{j=1}^{N} A_{ij} (v_j-v_i) +g(u_i,v_i),
\end{aligned}
\qquad (i=1,2,\cdots, N),
\end{equation}
where 
$$
A_{ij}=
\begin{cases}
1 & \text{if the $i$th and the $j$th vertices are connected},\\
0 & \text{if disconnected,} 
\end{cases}
$$
and $A_{ii}=0$ because we do not consider any self-loop in the present paper. Moreover, since we focus on undirected graphs, the matrix $A$ with the elements $A_{ij}$ is a symmetric matrix with $A_{ji}=A_{ij}$. And, the positive constants $d_u$ and $d_v$ mean respectively diffusivities of these quantities between vertices.  Also, the number of edges connecting to the $i$th vertex is expressed as $k_i: =\sum_{j=1}^{N} A_{ij}$. Thus, for each $i\in \{1,2,\cdots, N\}$, we can rewrite the flux term as 
$$
\sum_{j=1}^{N} A_{ij} (u_j-u_i)=-\sum_{j=1}^{N} L_{ij} u_j, 
$$
where 
$L_{ij}=\delta_{ij}k_i-A_{ij}$ ($\delta_{ij}$ is Kronecker's delta). The matrix $L$ with the elements $L_{ij}$ is called Laplacian matrix of the graph $A$. We note that the Laplacian matrix $L$ varies according to network topology, and eigenvalues of the Laplacian matrix $L$ give an important information on the Turing instability of a homogeneous steady state on the graph. (See the Methods section in \cite{NM}.) Here, we define a homogeneous steady state on a graph as $(u_i,v_i)=(\overline{u},\overline{v})$ for all $i$ in \eqref{NRD}. Our purpose is to investigate the stability of the homogeneous steady state. To this end, it suffices to consider the linearized system around the steady state. Substituting $(u_i,v_i)=(\overline{u}+\tilde{u}_i, \overline{v}+\tilde{v}_i)$ into \eqref{NRD} and neglecting higher order terms, we obtain the following linear reaction-diffusion model on the graph: 
\begin{equation}
\label{NRD2}
\begin{aligned}
\frac{du_i}{dt}&= -d_u\sum_{j=1}^{N} L_{ij} u_j+a u_i+b v_i,\\
\frac{dv_i}{dt}&= -d_v\sum_{j=1}^{N} L_{ij} v_j +c u_i+d v_i, 
\end{aligned}
\qquad (i=1,2,\cdots,N), 
\end{equation}
where we denoted $(\tilde{u}_i,\tilde{v}_i)$ by $(u_i,v_i)$ again. We call \eqref{NRD2} a reaction-diffusion model on a graph in this paper. 
One can write \eqref{NRD2} as 
\begin{equation}
\label{NRD3}
\begin{aligned}
\frac{d \mathbf{u}}{dt}&= -d_u L \mathbf{u}+a \mathbf{u}+b \mathbf{v},\\
\frac{d \mathbf{v}}{dt}&= -d_v L \mathbf{v} +c \mathbf{u}+d \mathbf{v}, 
\end{aligned}
\end{equation}
where $\mathbf{u}=(u_1,\cdots, u_N)$ and $\mathbf{v}=(v_1,\cdots, v_N)$. 
Obviously, we know that \eqref{NRD2} possesses the homogeneous steady state $(\mathbf{u},\mathbf{v})=(\mathbf{0},\mathbf{0})$, and we study the stability of the steady state. 
Here, we note that when we impose the condition \eqref{parameter stable} on the parameters and $d_u=d_v=0$, $(\mathbf{u},\mathbf{v})=(\mathbf{0},\mathbf{0})$ is stable. 
Therefore, we are interested to know how the stability of $(\mathbf{u},\mathbf{v})=(\mathbf{0},\mathbf{0})$ changes according to network topology and the values of $d_u$ and $d_v$. 
In this paper, we use the following parameter values 
$$
\begin{pmatrix}
a & b\\
c & d
\end{pmatrix}
=
\begin{pmatrix}
1 & -2\\
2 & -2
\end{pmatrix}
$$
in all numerics, which satisfies the condition \eqref{parameter stable}. 
In order to specify the Laplacian matrix $L$ in \eqref{NRD3}, different types of models are proposed. Below, we investigate a relation between the Turing instability and network topology which is generated by the various models. 
Let $\lambda_i$ and ${\boldsymbol \phi}_i$ $(i=1,\cdots, N)$ be eigenvalues of the matrix $L$ and the associated eigenvectors to them, i.e. $L{\boldsymbol \phi}_i=\lambda_i {\boldsymbol \phi}_i$ $(i=1,\cdots, N)$. 
Since \eqref{NRD3} is linear, we can express the solution as 
$\mathbf{u}(t)=\sum_{i=1}^N \alpha_i e^{\rho_i t}{\boldsymbol \phi}_i$ and $\mathbf{v}(t)=\sum_{i=1}^N \beta_i e^{\rho_i t}{\boldsymbol \phi}_i$, 
where $\rho_i$ and $(\alpha_i,\beta_i)$ are eigenvalues and their associated eigenvectors of the matrix 
$$
\begin{pmatrix}
-d_u \lambda_i +a & b\\
c & -d_v\lambda_i+d
\end{pmatrix}
\quad (i=1,\cdots,N). 
$$
Therefore, in order to check the stability of $(\mathbf{u},\mathbf{v})=(\mathbf{0},\mathbf{0})$, the real part of $\rho_i$ is important. 
If the real parts of all eigenvalues $\rho_i$ $(i=1,\cdots, N)$ are negative, then the homogeneous steady state $(\mathbf{u},\mathbf{v})=(\mathbf{0},\mathbf{0})$ is stable because $(\mathbf{u}(t),\mathbf{v}(t))=(\sum_{i=1}^N \alpha_i e^{\rho_i t}{\boldsymbol \phi}_i, \sum_{i=1}^N \beta_i e^{\rho_i t}{\boldsymbol \phi}_i)\to (\mathbf{0},\mathbf{0})$ as $t\to \infty$. If real part of at least one eigenvalue $\rho_j$ is positive, we have $e^{\rho_j t}\to \infty$ as $t\to \infty$, so the steady state $(\mathbf{u},\mathbf{v})=(\mathbf{0},\mathbf{0})$ is unstable. 
As a result, we find out the stability of the steady state $(\mathbf{u},\mathbf{v})=(\mathbf{0},\mathbf{0})$ by investigating the sign of real part of the eigenvalues $\rho_i$. 
To do this, 
we introduce some properties of eigenvalues of the Laplacian matrix of a graph. Assume that eigenvalues are sorted in ascending order, namely $\lambda_1\le \lambda_2\le \lambda_3\le \cdots$. 
The number of eigenvalues (with multiplicity) is $N$, which is the same as the number of vertices, and all the eigenvalues are real and non-negative. 
Since we deal with only connected graphs in the present paper, the smallest eigenvalue $\lambda_1$ is zero, which is simple, and the second smallest eigenvalue $\lambda_2$ is greater than zero. 
From linear algebra, we know that $\sum_{i=1}^{N}\lambda_i=\mathrm{tr}L=\sum_{i=1}^{N}k_i=2M$, where $M$ is the total number of edges. 
Therefore, the largest eigenvalue is roughly estimated by $\lambda_N<2M$.

Under the condition \eqref{parameter stable}, we discuss the sign of eigenvalues $\rho_i$ $(i=1,\cdots,N)$. 
From simple calculation, if both $a$ and $d$ are negative, all the eigenvalues $\rho_i$ $(i=1,\cdots, N)$ are negative, so that the instability of the homogeneous steady state does not occur. 
In order to yield the instability, either case $d<0<a$ or $a<0<d$ is necessary. 
Without loss of generality, we here assume the case 
\begin{equation}
\label{parameter stable 2}
d<0<a.
\end{equation}
For each $i$, the bifurcation curve where an eigenvalue $\rho_i$ takes zero is given on $(d_u,d_v)$ plane as follows:  
$$
\Gamma_i=\{ (d_u,d_v)\in \mathbb{R}_+^2 \, |\, (d_u\lambda_i-a)(d_v\lambda_i-d)-bc=0 \}
\quad (i=1,2,\cdots,N),
$$
where $\lambda_i$ is the $i$th eigenvalue of the Laplacian matrix of a graph. 
When we view $d_v$ as the function of $d_u$ for each curve $\Gamma_i$, the asymptote of the curve is $d_u=\frac{a}{\lambda_i}$. 
From information on the bifurcation curves, we can indicate the unstable region of the homogeneous steady state $(\mathbf{u},\mathbf{v})=(\mathbf{0},\mathbf{0})$ as $\cup_{i=1}^{N}D_i$, where $D_i=\{ (d_u,d_v)\in \mathbb{R}_+^2 \, | \, (d_u\lambda_i-a)(d_v\lambda_i-d)-bc<0 \}$. In addition, we remark that the size of each eigenvalue $\lambda_i$ of the Laplacian matrix $L$ determines properties of the corresponding bifurcation curve. 
\begin{figure}[htbp]
\begin{center}
\includegraphics[width=60mm]{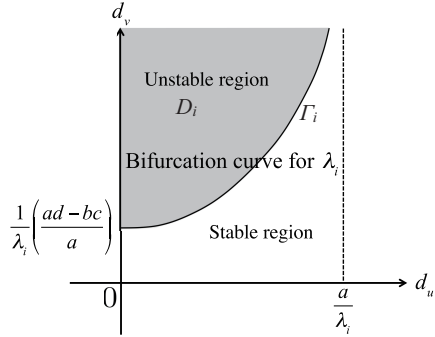}
\caption{Bifurcation curve and unstable region for the $i$th eigenvalue $\lambda_i$. }
\label{figure_BC}
\end{center}
\end{figure}
As shown in Figure \ref{figure_BC}, the bifurcation curve for $\lambda_i$ intersects with the $d_v$-axis at $d_v=\frac{1}{\lambda_i}\left( \frac{ad-bc}{a} \right)$ and asymptotically approaches the line $d_u=\frac{a}{\lambda_i}$. 
That is, if $\lambda_i$ is small, the value of the intersection point $\frac{1}{\lambda_i}\left(\frac{ad-bc}{a}\right)$ is large and the asymptomatic value $\frac{a}{\lambda_i}$ is also large. 
On the other hand, when $\lambda_i$ is large, both $\frac{1}{\lambda_i}\left( \frac{ad-bc}{a} \right)$ and $\frac{a}{\lambda_i}$ are small. 
Therefore, we find out that the distribution of eigenvalues plays an important role in determining stable and unstable regions on $(d_u,d_v)$ plane of the homogeneous steady state of \eqref{NRD2}. 
In particular, we state that small eigenvalues determine the unstable region for large $(d_u,d_v)$ and large eigenvalues contribute to the determination of the unstable region for small $(d_u,d_v)$. 
After this point, the stable and unstable regions of the homogeneous steady state on $(d_u,d_v)$ plane is briefly called stable-unstable region.

In this paper, we consider three models which generate graphs, i.e. the Erd\H{o}s-R\'enyi, the Watts-Strogatz, and the threshold network models. 
Each model stochastically produces a graph.

\paragraph{Erd\H{o}s-R\'enyi model}
\begin{itemize}
\item An edge is set between each pair of distinct vertices with probability $p$, independently of the other vertices.
\end{itemize}

\paragraph{Watts-Strogatz model}
\begin{enumerate}
\item Let $B$ be an even integer. We prepare a regular enhanced ring which is defined by the adjacency matrix $A_{ij}=1$ if $0<j-i+N\leq \frac{B}{2}\quad \mbox{or}\quad 0<i-j+N\leq \frac{B}{2} \mod N$, otherwise $A_{ij}=0$. We define the edge set of the regular enhanced ring as $E=\{(i,j)\, |\, A_{ij}=1\}$, and set $E^{\prime }=E$ initially. 
\item For each edge $(i,j)\in E$, we perform the next procedure with probability $p$. We choose one of two vertices $i,j$ with probability $1/2$ and set the chosen vertex as $I$. We uniformly select a vertex $k$ such that $(I,k)\notin E^{\prime}\cup E\cup (I,I)$. Then we reset $E^{\prime }$ to be $\left( E^{\prime }\setminus (i,j) \right) \cup (I,k)$.
\item The resulting network with the edge set $ E^{\prime }$ is a network generated by the Watts-Strogatz model.
\end{enumerate}

\paragraph{Threshold network model}
\begin{enumerate}
\item For $i=1,2,\ldots,N$, a weight $w_i$ is randomly chosen by an exponential distribution with a parameter $\Lambda$. 
\item Given a value $\theta$ which is called threshold value, for every pair of distinct vertices $i$ and $j$, if the sum of weights $w_i+w_j$ is greater than the threshold value $\theta$, the two vertices are connected by an edge. 
\end{enumerate}

Our interest is how network topology influences the Turing instability, but graphs are stochastically generated by the models. 
In addition, as we agued above, eigenvalues of the Laplacian matrix of a graph play an important role in Turing instability on a graph. 
Since a graph is stochastically generated, the distribution of the eigenvalues also changes according to the stochastic network construction. 
Therefore, we represent stable and unstable regions of the homogeneous steady state with probability by repeating the following trial for $500$ samples of the graph and taking the average: 
\begin{enumerate}
\item calculate eigenvalues of the Laplacian matrix for a network generated by a model. 
\item compute the stable and unstable regions of $(\mathbf{u},\mathbf{v})=(\mathbf{0},\mathbf{0})$ in $(d_u,d_v)$ plane, based on the bifurcation curves. 
\item represent the unstable region as one and the stable region as zero on each lattice point in $(d_u,d_v)$ plane. 
\end{enumerate}
Consequently, we present the probability that the homogeneous steady state $(\mathbf{u},\mathbf{v})=(\mathbf{0},\mathbf{0})$ is destabilized in $(d_u,d_v)$ plane. 

\begin{figure}[!h]
\begin{center}
\includegraphics[width=120mm]{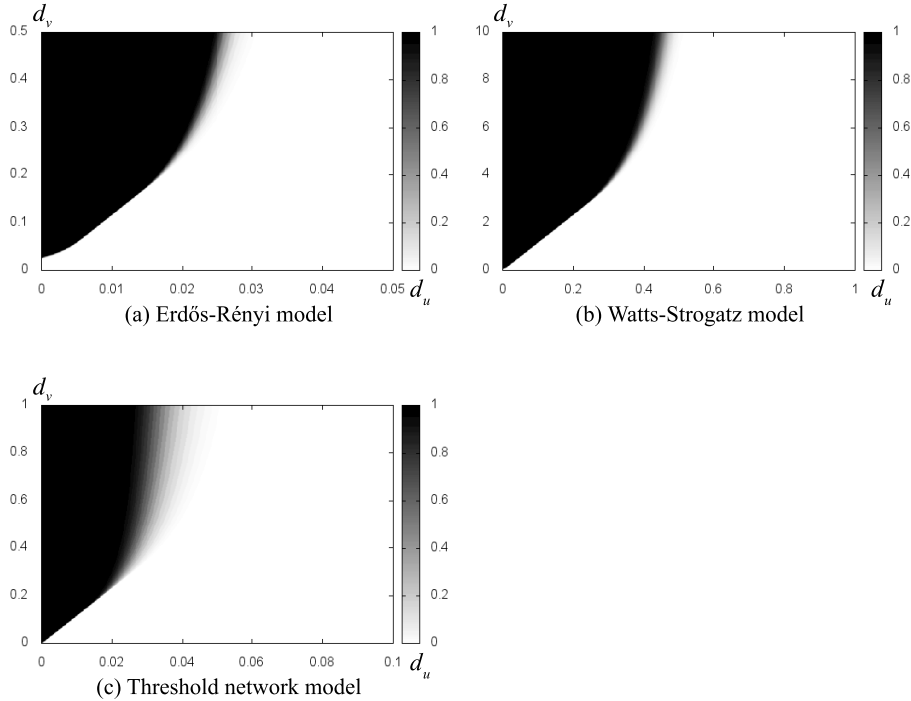}
\caption{The stable-unstable regions of the homogeneous steady state $(\mathbf{u},\mathbf{v})=(\mathbf{0},\mathbf{0})$ for \eqref{NRD2} with $N=500$. The horizontal and vertical axes indicate $d_u$ and $d_v$, respectively. 
(a) The Erd\H{o}s-R\'enyi model with $p=0.1$. 
(b) The Watts-Strogatz model with $B=20$ and $p=0.1$. 
(c) The threshold network model with $\Lambda=1$ and $\theta=3$. }
\label{figure2}
\end{center}
\end{figure}

Figure \ref{figure2} shows the stable-unstable regions of the steady state $(\mathbf{u},\mathbf{v})=(\mathbf{0},\mathbf{0})$ in \eqref{NRD2} on graphs with 500 vertices generated by the Erd\H{o}s-R\'enyi, the Watts-Strogatz and the threshold network models, respectively. 
These figures mean that the black region denotes that the steady state $(\mathbf{u},\mathbf{v})=(\mathbf{0},\mathbf{0})$ is unstable with probability one and the white one denotes that it is unstable with probability zero, that is, stable. 
We express the probability between those with tones of gray. 
It appears that the stable-unstable regions do not differ so much each other though the scale of the axes is different. 
However, we will see that the parameters which determine the probability in the models and the number of vertices $N$ strongly influence the stable-unstable regions.

%%%%%%%%%%%%%%%%%%%%%%%%%%%%%%%%%%%%%%%%%%

\section{Networks generated by the Erd\H{o}s-R\'enyi model}
\label{Networks generated by Erdos-Renyi model}

\begin{figure}[!h]
\begin{center}
\includegraphics[width=120mm]{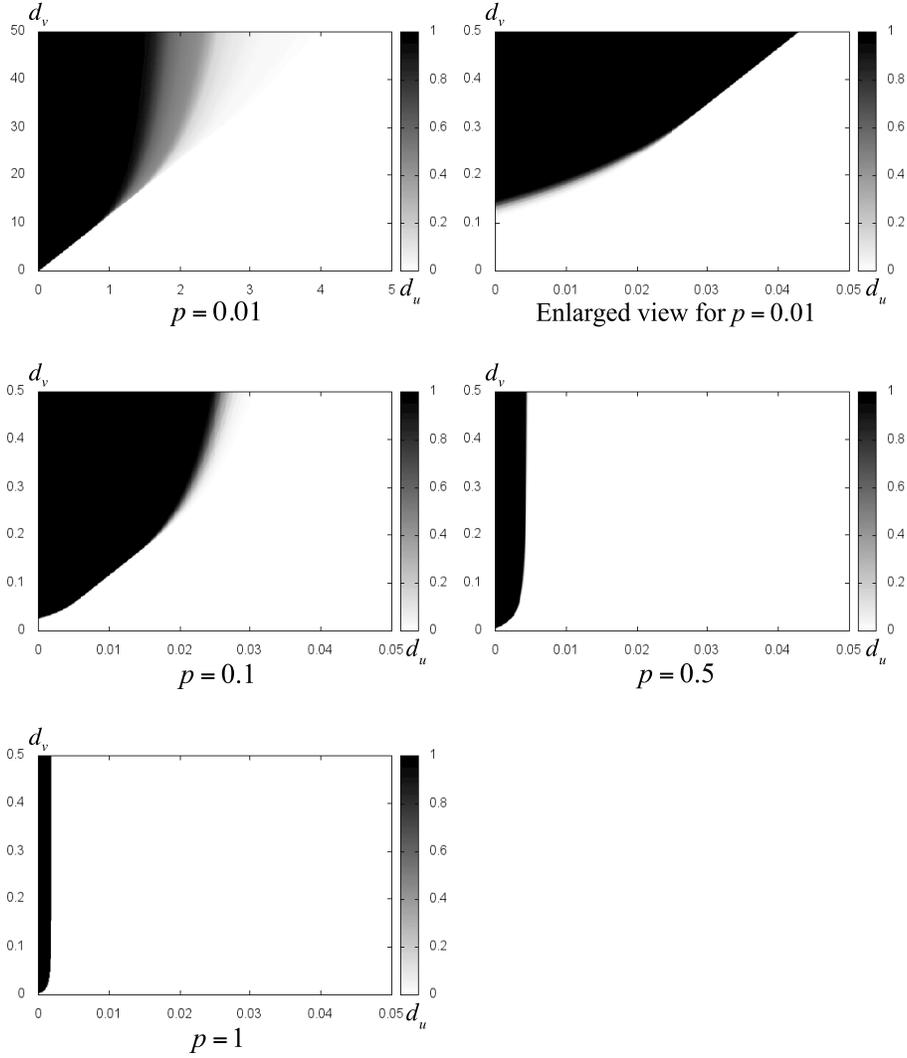}
\caption{The stable-unstable regions of the homogeneous steady state $(\mathbf{u},\mathbf{v})=(\mathbf{0},\mathbf{0})$ for \eqref{NRD2} on the Erd\H{o}s-R\'enyi random graphs. The number of vertices is $N=500$ and the link probability between vertices $p$ is indicated at the bottom of figures. The black and white regions denote to be unstable with probability one and unstable with probability zero, respectively. The horizontal and vertical axes indicate $d_u$ and $d_v$, respectively. }
\label{SUR_ER}
\end{center}
\end{figure}

In this section, we deal with a reaction-diffusion model on a graph which is stochastically generated by the Erd\H{o}s-R\'enyi model and reveal the stable-unstable region of the homogeneous steady state $(\mathbf{u},\mathbf{v})=(\mathbf{0},\mathbf{0})$ of \eqref{NRD2} from the linear stability analysis with the aid of a computer. 
Obviously, the number of edges is zero when the probability $p=0$, so this graph is disconnected. On the other hand, when $p=1$, it becomes a complete graph with $N$ vertices since every pair of distinct vertices is connected by an edge. Since the average degree of an Erd\H{o}s-R\'enyi random graph is $p(N-1)$, the closer the value $p$ approaches one or the larger the number of vertices becomes, the more edges a graph possesses. 
We are interested in the transition of the stable-unstable regions of the steady state $(\mathbf{u},\mathbf{v})=(\mathbf{0},\mathbf{0})$ for \eqref{NRD2}, depending on the link probability $p$ and the number of vertices $N$. 
We illustrate the stable-unstable regions of the homogeneous steady state of \eqref{NRD2} based on eigenvalues of the Laplacian matrix $L$. 
Consequently, we present the probability that the steady state $(\mathbf{u},\mathbf{v})=(\mathbf{0},\mathbf{0})$ is destabilized, as shown in Figure \ref{SUR_ER} when $N=500$ and the value of $p$ is varied. 
From these results, we know that the region where the homogeneous steady state is unstable with high probability is large when $p$ is small ($p=0.01$), while the unstable region gradually shrinks as the value of $p$ approaches one. 
In general, for the fully connected graph with $N$ vertices, eigenvalues of the Laplacian matrix $L$ are $\lambda_1=0$ and $\lambda_i=N$ $(i=2,\cdots, N)$. Therefore, the stable-unstable region for $p=1$ and $N=500$ is deterministically obtained. 
On the other hand, if $(d_u,d_v)$ takes a pair of positive values in the vicinity of the origin, for instance $(d_u,d_v)=(0.005,0.1)$, the steady state $(\mathbf{u},\mathbf{v})=(\mathbf{0},\mathbf{0})$ is stable with high probability when $p=0.01$, but it gets to be contained in the unstable region as the value $p$ increases ($p=0.1$). We find that there exists a certain parameter region on $(d_u, d_v)$ plane such that the stability of $(\mathbf{u},\mathbf{v})=(\mathbf{0},\mathbf{0})$ changes from the stable state to the unstable one when the value of $p$ increases. Interestingly, this means that the destabilization occurs when the connectivity between vertices rises, that is, when the number of edges increases. This network connectivity induced instability is a quite new phenomenon. When we increase the value $p$ further ($p=0.5$ and $1$), the homogeneous steady state becomes stable with high probability again for $(d_u,d_v)=(0.005,0.1)$. This result implies that the network connectivity induced instability occurs when a pair $(d_u,d_v)$ and the number of edges are suitable. 
\begin{figure}[!h]
\begin{center}
\includegraphics[width=120mm]{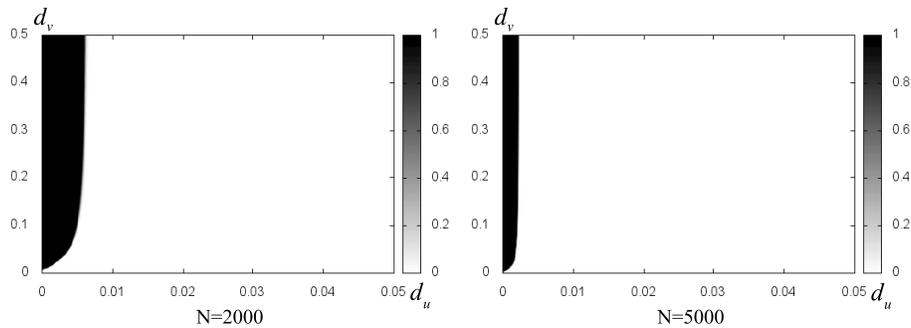}
\caption{The stable-unstable regions of the homogeneous steady state $(\mathbf{u},\mathbf{v})=(\mathbf{0},\mathbf{0})$ for \eqref{NRD2} on the Erd\H{o}s-R\'enyi random graphs. The number of vertices is $N=2000$ and $N=5000$, and the connection probability between vertices is $p=0.1$. The black and white regions denote to be unstable with probability one and unstable with probability zero, respectively. The horizontal and vertical axes indicate $d_u$ and $d_v$, respectively. }
\label{figure4}
\end{center}
\end{figure}
Next, when the value $p$ is fixed and the number of vertices $N$ is varied, we observe the dependence of the stable-unstable regions on $N$. In the Erd\H{o}s-R\'enyi model with a fixed $p$, when the number of vertices increases, the average degree also increases. Roughly, this means that the destabilization of the homogeneous steady state $(\mathbf{u},\mathbf{v})=(\mathbf{0},\mathbf{0})$ may be inhibited because the active dispersal of substances between vertices which comes from a lot of connections smoothes the heterogeneity. 
Actually, this expectation is fit for the numerical results in Figures \ref{SUR_ER} and \ref{figure4}. For $p=0.1$, we find that the unstable region becomes narrow, depending on an increase of the number of vertices. 
These transitions of the stable-unstable regions can be understood by the behavior of eigenvalues $\lambda_i$ $(i=1,2,3,\cdots)$. 
Generally, in the Erd\H{o}s-R\'enyi random graphs, the largest eigenvalue of the Laplacian matrix is approximately $\lambda_N\approx Np$ when $N\gg 1$ (\cite{DJ}). This means that the largest eigenvalues $\lambda_N$ becomes large according to increases of the values $N$ and $p$. And, we numerically checked that small eigenvalues except for $\lambda_1$ also become large as increases in the values $N$ and $p$, so that the unstable region shrinks. (See also Figure \ref{figure_BC}.)
However, when we change both the values $N$ and $p$ while keeping the average degree, eigenvalue distributions are quite similar, therefore, the stable-unstable regions of the homogeneous steady state hardly change, as shown in Figure \ref{eigenvalue_ER4}. 
\begin{figure}[!h]
\begin{center}
\includegraphics[width=120mm]{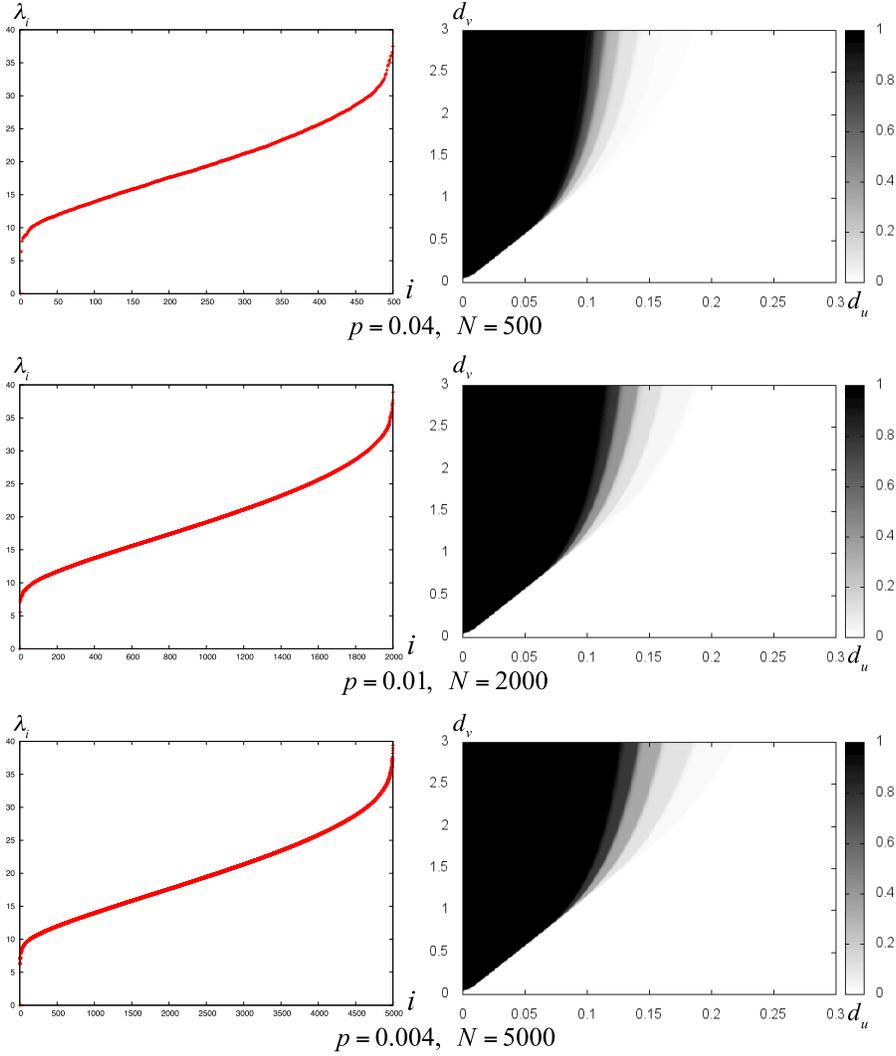}
\caption{(Left) Examples of all eigenvalues $\{ \lambda_i : 1\le i\le N\}$ of the Laplacian matrix for \eqref{NRD2} on an Erd\H{o}s-R\'enyi random graph with the almost equal average degree $p(N-1)$. The horizontal and vertical axes mean eigenvalue numbers $i$ and these values $\lambda_i$, respectively. Eigenvalues are sorted in ascending order, that is $0=\lambda_1<\lambda_2\le\cdots \le \lambda_{N}$. (Right) The stable-unstable regions of the steady state $(\mathbf{u},\mathbf{v})=(\mathbf{0},\mathbf{0})$ for \eqref{NRD2} on the Erd\H{o}s-R\'enyi random graphs. The horizontal and vertical axes indicate $d_u$ and $d_v$, respectively. The number of vertices $N$ and the link probability $p$ are indicated at the bottom of each figure. }
\label{eigenvalue_ER4}
\end{center}
\end{figure}
Thus, in the Erd\H{o}s-R\'enyi model, we can suggest that the stable-unstable region of the homogeneous steady state changes, depending strongly on the average degree. 
In other words, if the average degree of two Erd\H{o}s-R\'enyi random graphs is almost equal, the stable-unstable regions resemble each other.

%%%%%%%%%%%%%%%%%%%%%%%%%%%%%%%%%%%%%%%%%%

\section{Networks generated by the Watts-Strogatz model}
\label{Networks generated by Watts-Strogatz model}

\begin{figure}[!h]
\begin{center}
\includegraphics[width=120mm]{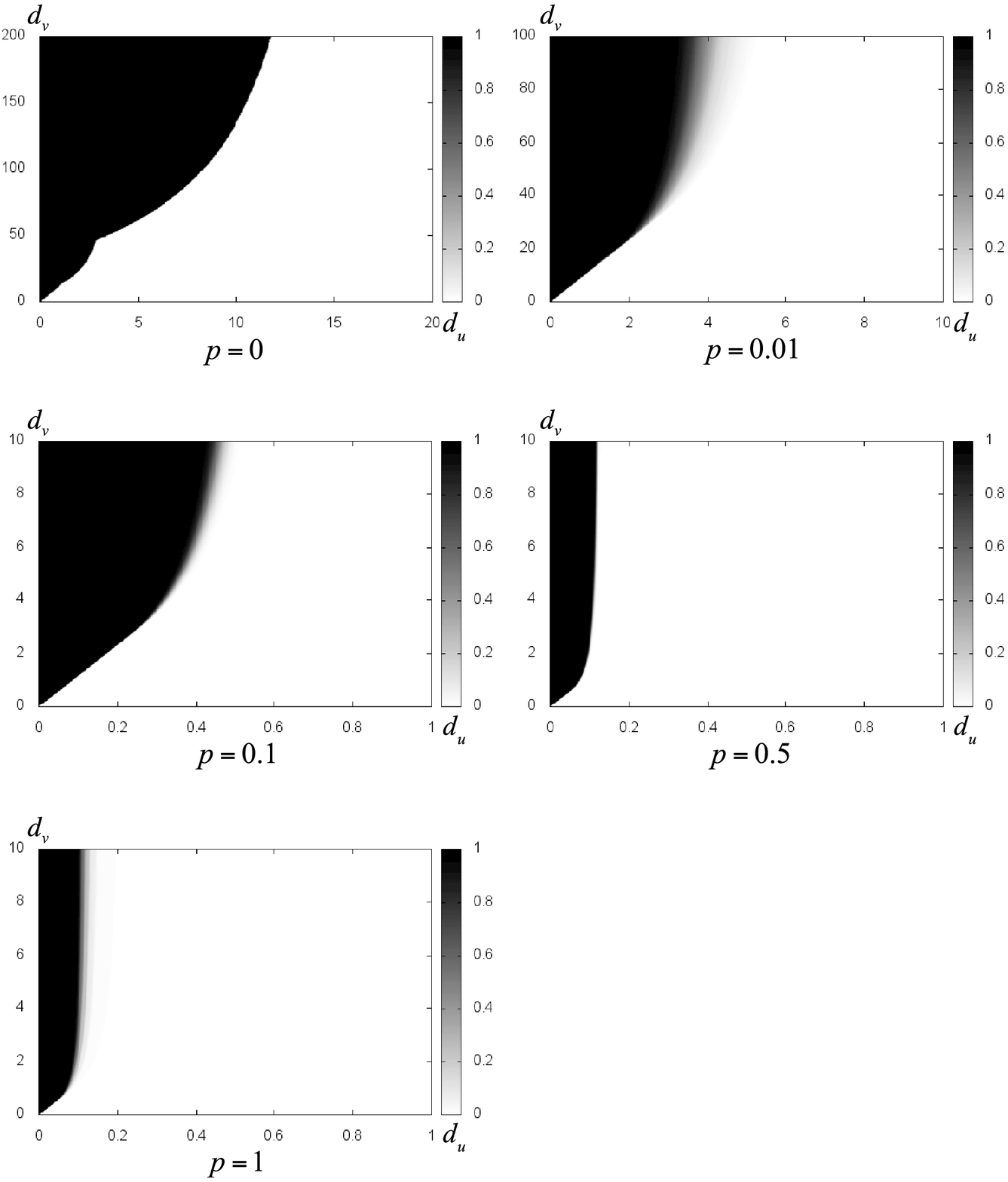}
\caption{The stable-unstable regions of the homogeneous steady state $(\mathbf{u},\mathbf{v})=(\mathbf{0},\mathbf{0})$ in a reaction-diffusion model on networks generated by the Watts-Strogatz model. The number of vertices is $N=500$, the average degree is $B=20$, and the reconnection probability $p$ is indicated at the bottom of each figure. The black and white regions denote to be unstable with probability one and unstable with probability zero, respectively. The horizontal and vertical axes indicate $d_u$ and $d_v$, respectively. }
\label{SUR_WS}
\end{center}
\end{figure}

\begin{figure}[!h]
\begin{center}
\includegraphics[width=120mm]{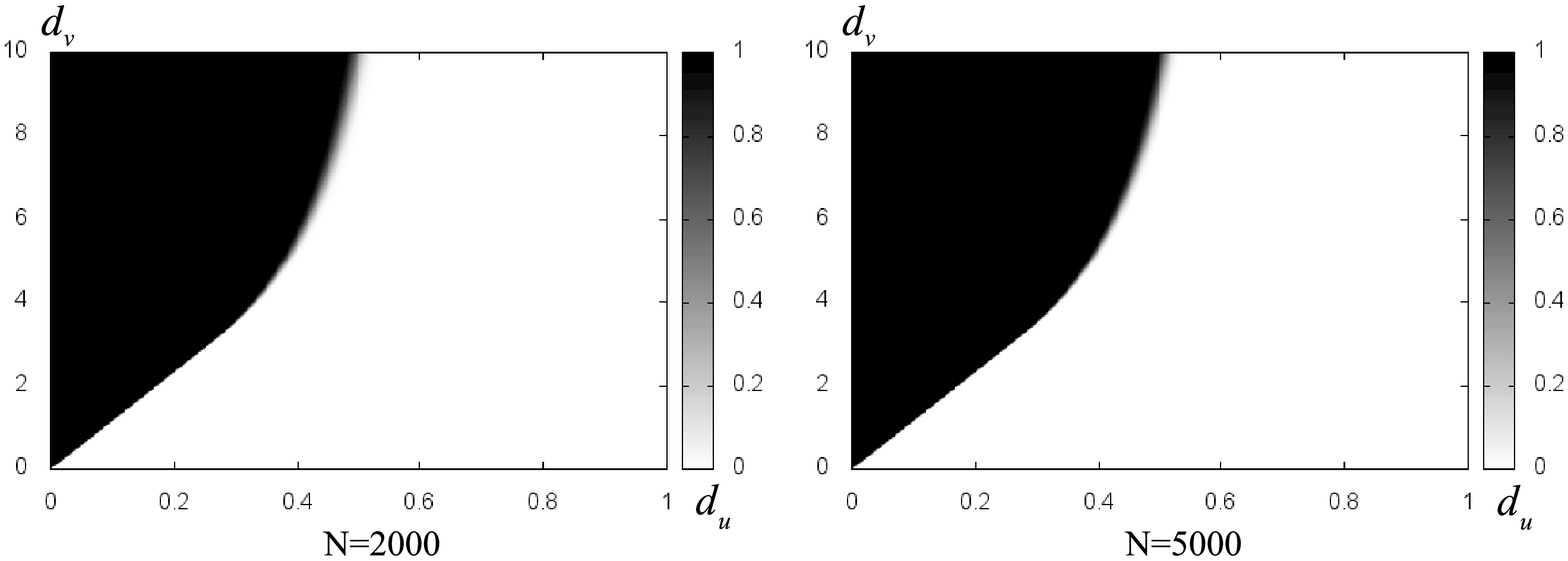}
\caption{The stable-unstable regions of the homogeneous steady state $(\mathbf{u},\mathbf{v})=(\mathbf{0},\mathbf{0})$ in a reaction-diffusion model on networks generated by the Watts-Strogatz model. The number of vertices is $N=2000$ and $N=5000$, the average degree is $B=20$, and the reconnection probability is $p=0.1$ . The black and white regions denote to be unstable with probability one and unstable with probability zero, respectively. The horizontal and vertical axes indicate $d_u$ and $d_v$, respectively. }
\label{figure7}
\end{center}
\end{figure}

\begin{figure}[!h]
\begin{center}
\includegraphics[width=120mm]{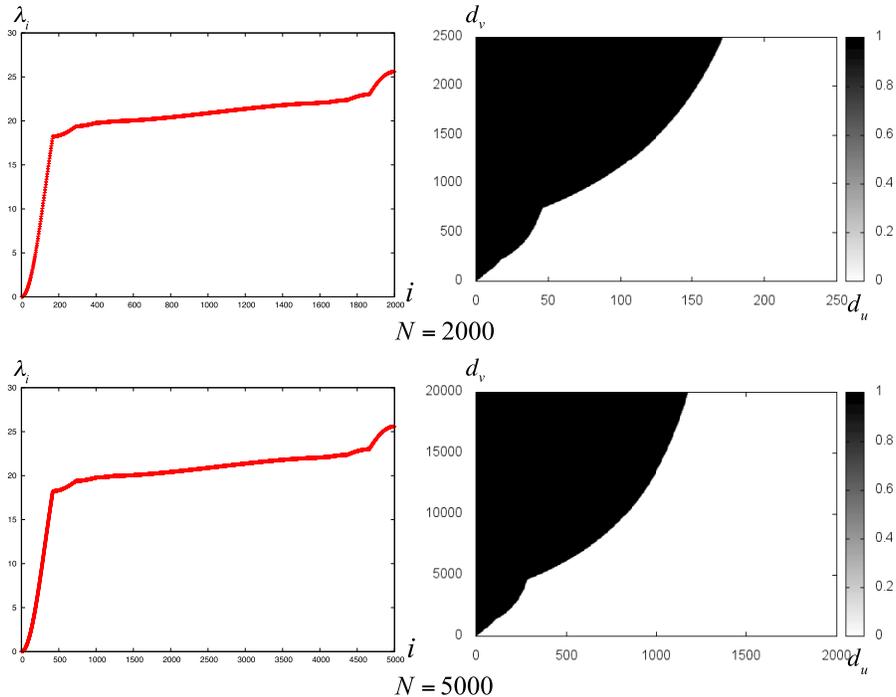}
\caption{(Left) Examples of all eigenvalues $\{\lambda_i:1\le i\le N\}$ of the Laplacian matrix for \eqref{NRD2} on a regular ring lattice. The horizontal and vertical axes mean eigenvalue numbers $i$ and these values $\lambda_i$, respectively. (Right) The stable-unstable regions of the homogeneous steady state $(\mathbf{u},\mathbf{v})=(\mathbf{0},\mathbf{0})$ in a reaction-diffusion model on networks generated by the Watts-Strogatz model. The number of vertices is $N=2000$ and $N=5000$, the average degree is $B=20$, and the reconnection probability is $p=0$ . The black and white regions denote to be unstable with probability one and unstable with probability zero, respectively. The horizontal and vertical axes indicate $d_u$ and $d_v$, respectively. }
\label{figure8}
\end{center}
\end{figure}

We deal with a reaction-diffusion model on a complex network generated by the Watts-Strogatz model in this section. 
First, we briefly explain the properties of graphs generated by the Watts-Strogatz model, that is,  network topology completely differs, depending on the rewiring probability $p$. For example, when $p=0$, the network remains to be the regular ring lattice with the degree $B$, and when $p=1$, it possesses similar properties to an Erd\H{o}s-R\'enyi random graph because all edges are randomly reconnected to other vertices though there is a restriction that an edge is never rewired within a vertex of $B$ neighbors. We are interested in the transition of the stable-unstable region of the homogeneous steady state, depending on the values of $p$ (the rewiring probability) and $N$ (the number of vertices). 
We illustrate the stable-unstable regions of the steady state $(\mathbf{u},\mathbf{v})=(\mathbf{0},\mathbf{0})$ of \eqref{NRD2}, as shown in Figures \ref{SUR_WS}, \ref{figure7} and \ref{figure8}. 
In this case also, since a network is constructed stochastically, we indicate the stable-unstable regions with probability by performing 500 trials and taking the average as well as the case of the Erd\H{o}s-R\'enyi model. 
Figure \ref{SUR_WS} shows the stable-unstable regions for $N=500$ when the value of $p$ varies. 
Interestingly, one can see from the figures that the unstable region of the homogeneous steady state becomes narrow as the value of $p$ increases though the average degree is invariant. 
This means that the destabilization of the steady state $(\mathbf{u},\mathbf{v})=(\mathbf{0},\mathbf{0})$ of \eqref{NRD2} tends to be inhibited as a regular ring lattice changes into a random graph. 
Since the networks constructed by the Watts-Strogatz model shortens the distance between any two vertices, that facilitates the dispersal of the quantities to all the vertices. 
Therefore, it seems that this is related to the shift in the unstable region.
Moreover, the stable-unstable region for $p=1$ in Figure \ref{SUR_WS} is quite similar to that for $p=0.04$ and $N=500$ in Figure \ref{eigenvalue_ER4} because all edges are randomly rewired to other vertices by the Watts-Strogatz model when $p=1$. 
Note that the average degree of the two cases is almost equal. 
Next, we investigate the transition of the stable-unstable region when the values $p$ and $B$ are fixed and the number of vertices $N$ varies. 
For $p=1$, we observed that the region does not change even when $N$ varies, because the networks for $p=1$ possess the similar properties to Erd\H{o}s-R\'enyi random graphs with the average degree $B$. So, the stable-unstable regions for $N=2000$ and $5000$ are very similar to those in Figure \ref{eigenvalue_ER4}. 
Interestingly, even though the rewiring probability is $p=0.1$, the stable-unstable regions hardly change when the number of vertices $N$ increases. (See Figure \ref{SUR_WS} for $p=0.1$ and Figure \ref{figure7}.)
However, we remark that the unstable region for $p=0$ grows when the number of vertices $N$ increases. This phenomenon comes from the behavior of small eigenvalues when the number $N$ varies. 
The numerical results for $p=0$ in Figure \ref{figure8} show that all the eigenvalues are less than a certain constant $c$ and all the eigenvalues is distributed in the range $[0,c]$ even when $N$ varies. 
This means that the larger the number $N$ becomes, the more packed all the eigenvalues becomes in the range $[0,c]$. 
Therefore, small eigenvalues extend the unstable region in the case of a regular ring lattice. (See Figure \ref{figure_BC}.) 
When the number of vertices $N$ is large enough, we will prove the existence of the stable-unstable region of the homogeneous steady state $(\mathbf{u},\mathbf{v})=(\mathbf{0},\mathbf{0})$ for the regular ring lattice ($p=0$) in section \ref{Theoretical results}. In the case of the Erd\H{o}s-R\'enyi model, we concluded that the stable-unstable regions did not vary if the average degree of networks was almost equal. However, we find from the results of the Watts-Strogatz model that the stable-unstable regions vary considerably, depending on network topology (structural properties of networks) even when the average degree is completely equal.

%%%%%%%%%%%%%%%%%%%%%%%%%%%%%%%%%%%%%%%

\section{Networks generated by the threshold network model}
\label{Networks generated by Threshold model}

\begin{figure}[!h]
\begin{center}
\includegraphics[width=120mm]{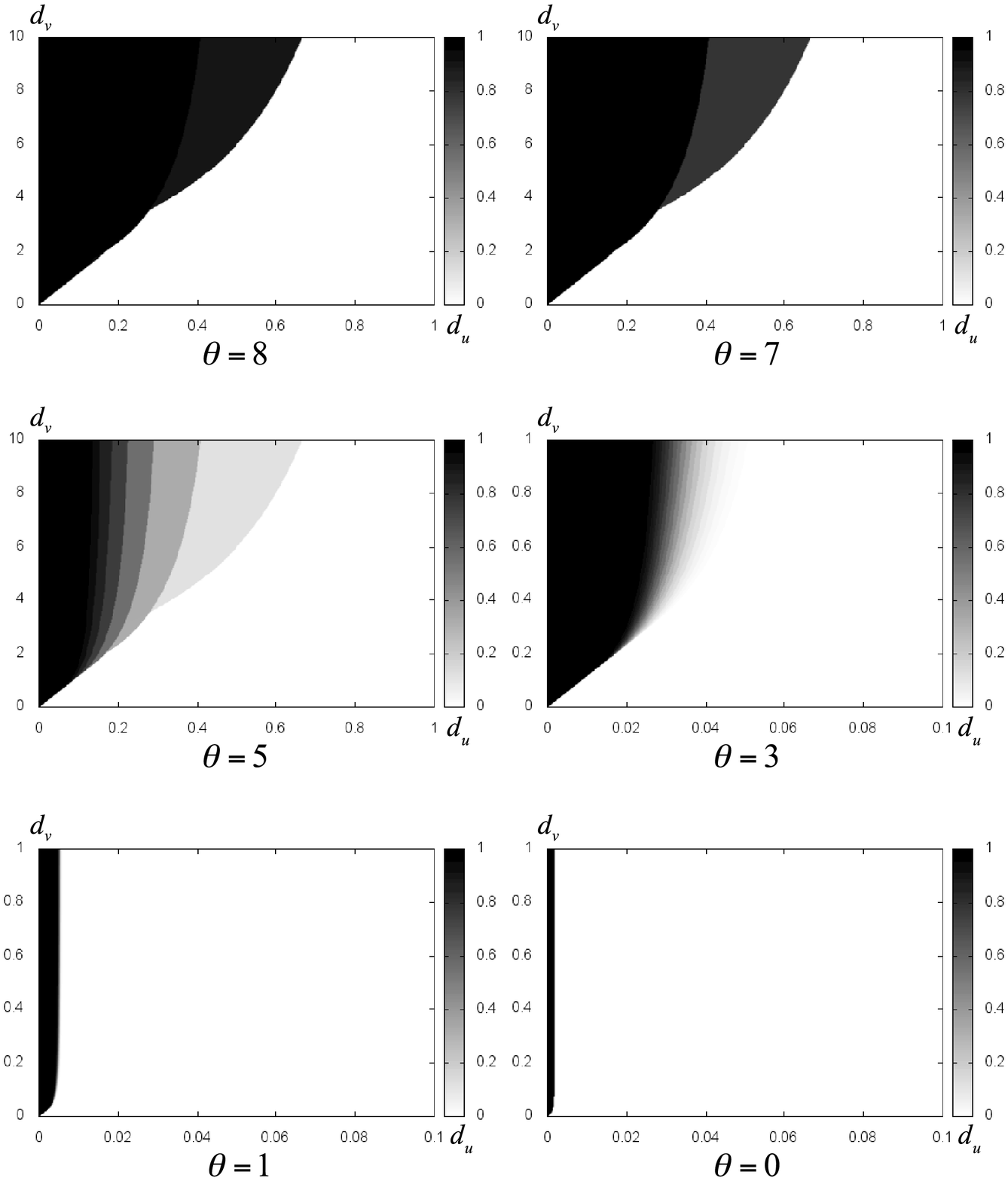}
\caption{The stable-unstable regions of the homogeneous steady state $(\mathbf{u},\mathbf{v})=(\mathbf{0},\mathbf{0})$ in a reaction-diffusion model on networks generated by the threshold network model. The number of vertices is $N=500$, the parameter of the exponential distribution $\Lambda=1$, and the threshold value $\theta$ is indicated at the bottom of each figure. The black and white regions denote to be unstable with probability one and unstable with probability zero, respectively. The horizontal and vertical axes indiate $d_u$ and $d_v$, respectively. }
\label{SUR_TV}
\end{center}
\end{figure}

\begin{figure}[!h]
\begin{center}
\includegraphics[width=120mm]{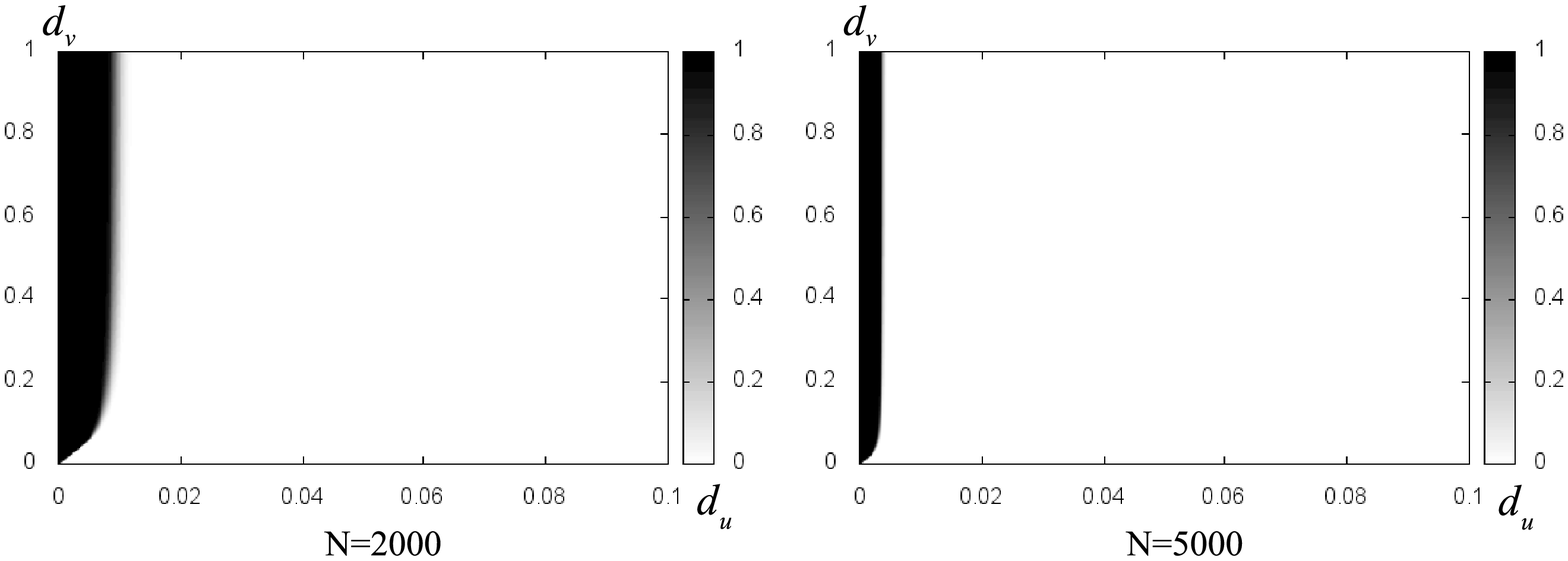}
\caption{The stable-unstable regions of the homogeneous steady state $(\mathbf{u},\mathbf{v})=(\mathbf{0},\mathbf{0})$ in a reaction-diffusion model on networks generated by the threshold network model. The number of vertices is $N=2000$ and $N=5000$, the parameter of the exponential distribution is $\Lambda=1$, and the threshold value is $\theta=3$. The black and white regions denote to be unstable with probability one and unstable with probability zero, respectively. The horizontal and vertical axes indicate $d_u$ and $d_v$, respectively. }
\label{figure10}
\end{center}
\end{figure}

As the third model which produces complex networks, we use the threshold network model and perform the linear stability analysis of a reaction-diffusion model on a network generated by it. It is well known that this model can produce scale free networks \cite{so, mmk04, scb}. 
From the construction by this model, we know that the generated graph is fully connected when $\theta=0$. On the other hand, it is said that this model can generate a scale-free network when the value of $\theta$ is appropriate, that is, the network possesses a small number of hubs. 
We note that the condition on connectedness of a graph produced by the threshold network model is $\min_{i\in\left\{1,2,\ldots,N\right\}}\{w_i\}+\max_{i\in\left\{1,2,\ldots,N\right\}}\{w_i\}\ge \theta$. 
Therefore, the vertex with the largest weight is connected to all of the other vertices. We illustrate the stable-unstable regions of the homogeneous steady state $(\mathbf{u},\mathbf{v})=(\mathbf{0},\mathbf{0})$ of \eqref{NRD2} when the values of $\theta$ and $N$ vary. As well as the discussion in sections \ref{Networks generated by Erdos-Renyi model} and \ref{Networks generated by Watts-Strogatz model}, we show it with probability by taking the average of $500$ trials. Figures \ref{SUR_TV} and \ref{figure10} show the stable-unstable regions of the steady state with probability for $N=500$, $2000$, and $5000$. 
When $\theta=0$, the graphs are fully connected. 
Therefore, the smallest eigenvalue is $\lambda_1=0$ and all of the other eigenvalues is $\lambda_i=N$ $(i=2,\cdots, N)$. 
As the value of $\theta$ is increased, we observe that the region where the homogeneous steady state is destabilized with high probability becomes wider for a fixed $N$. (See Figure \ref{SUR_TV}.)
This means that the unstable region expands as the number of edges diminishes, namely the decrease in the number of edges facilitates the destabilization of the homogeneous steady state as well as the case of the Erd\H{o}s-R\'enyi model. Next, we observe the transition of the stable-unstable region when the value $p$ is fixed and the number $N$ increases. For $\theta=0$, the distribution of eigenvalues is $\lambda_1=0$ and $\lambda_i=N$ $(i=2,\cdots, N)$, so that an increase in the number $N$ shrinks the stable-unstable regions. Figure \ref{SUR_TV} for $\theta=3$ and Figure \ref{figure10} show that this tendency also holds for $\theta \neq 0$, that is, the region where the homogeneous steady state is destabilized with high probability gradually shrinks as to an increase in the number $N$. 
When the value of $\theta$ is fixed and the number of vertices $N$ increases, the number of hubs in a network also increases. Since the hubs are connected to a lot of vertices, a flux of the quantities $u$ and $v$ between vertices promotes the smoothing of the quantities. Therefore, it seems that this is related to the occurrence of the phenomenon that the probability of the instability becomes low when the number of vertices increases. In the next section, we will theoretically refer to this phenomenon from the viewpoint of eigenvalue distribution of the Laplacian matrix of a graph generated by the threshold network model when the number of vertices is large enough.

%%%%%%%%%%%%%%%%%%%%%%%%%%%%%%%%%%%%%%%

\section{Approximate analysis}
\label{Theoretical results}
In this section, we prove some theoretical results on the Turing instability in a reaction-diffusion model on a graph. The first result is on the Turing instability on an enhanced graph, which we can also get from the Watts-Strogatz model with $p=0$ in section \ref{Networks generated by Watts-Strogatz model}. The second one is on eigenvalues of the Laplacian matrix of a graph generated by the threshold network model. 

\subsection{Turing instability on an enhanced cycle}
In this subsection, we deal with the stability of the homogeneous steady state $(\mathbf{u},\mathbf{v})=(\mathbf{0},\mathbf{0})$ on a diffusion process on an enhanced cycle when the number of its vertices is large enough.
We carry out the analysis under the condition $bc<0$.
Ahead of the analysis of the enhanced cycle, we make a brief discussion for a general graph $G$ with $N$ vertices.
Let $\mathcal{L}_{+}$ be the set $\mathcal{L}_{+}=\left\{l\in\left\{0,1,\ldots,N-1\right\}\,|\,\,g(\sigma_G(l))\geq 0\right\}$, where $\sigma_G(l)\,(l=0,1,\ldots,N-1)$ are the eigenvalues of the matrix $-L$ of the graph $G$ and 
\begin{equation*}
 g(s)=(d_u-d_v)^2 s^2+2(a-d)(d_u-d_v)s+(a-d)^2+4bc.
\end{equation*}
The function $g(s)$ comes from a discriminant of a characteristic equation of a $2\times 2$ matrix 
$$
\begin{pmatrix}
d_u s+a & b\\
c & d_v s+d
\end{pmatrix}.
$$
After straightforwardly analyzing the characteristic polynomial of the $2N\times 2N$ matrix
\begin{equation}
 \left[\left(-L_{ij}\left(\begin{array}{cc}
	d_u& 0\\
	      0 & d_v
	     \end{array}\right)
 +\delta_{ij}\left(\begin{array}{cc}
	      a& b\\
		    c& d
		   \end{array}\right)\right)_{ij}\right],
 \label{eq:TM:diffusuin matrix}
\end{equation}
we realize that the homogeneous steady state is stable if and only if
\begin{equation}
 \left\{\begin{array}{l}
  a+d <0,\\
	 \min_{l\in \mathcal{L}_{+}}\tau(\sigma_G(l)) >0,
	\end{array}\right.\label{eq:d_condition}
\end{equation}
where
\begin{equation*}
 \tau(s)=d_ud_vs^2+(d d_u+ad_v)s+ad-bc.
\end{equation*}
We suppose $\min_{l\in \mathcal{L}_{+}}\tau(\sigma_G(l)) >0$ for the null set $\mathcal{L}_{+}=\phi$.
Note that (\ref{NRD2}) is rewritten as
\begin{equation*}
 \frac{d}{dt}\left(\begin{array}{c}
			     u_i\\ v_i
				  \end{array}\right)
 =\sum_{j=1}^N 
 \left\{-L_{ij}\left(\begin{array}{cc}
  d_u& 0\\
	0 & d_v
       \end{array}\right)
 +\delta_{ij}\left(\begin{array}{cc}
	      a& b\\
		    c& d
		   \end{array}\right)\right\}
 \left(\begin{array}{c}
  u_j\\ v_j
       \end{array}\right).
 \label{eq:TM:NRD2}
\end{equation*}

\begin{figure}[!h]
 \begin{center}
  \begin{minipage}{35mm}
   \begin{center}
    \includegraphics[scale=0.3]{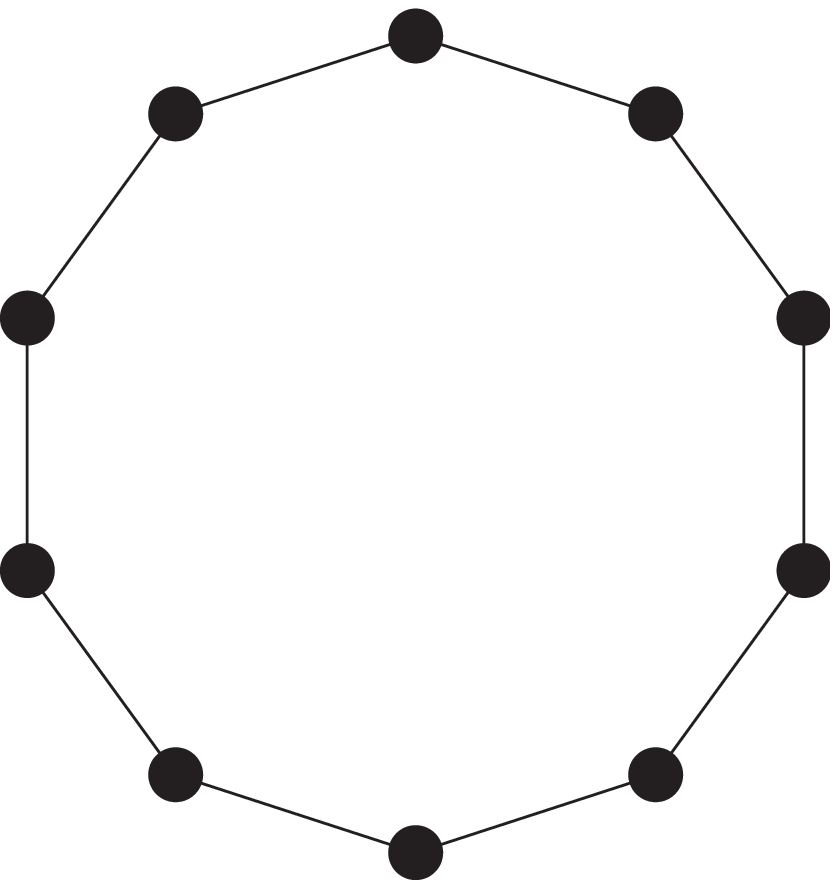}\\
    {(a) $k=1$}
   \end{center}
  \end{minipage}
   \begin{minipage}{35mm}
   \begin{center}
    \includegraphics[scale=0.3]{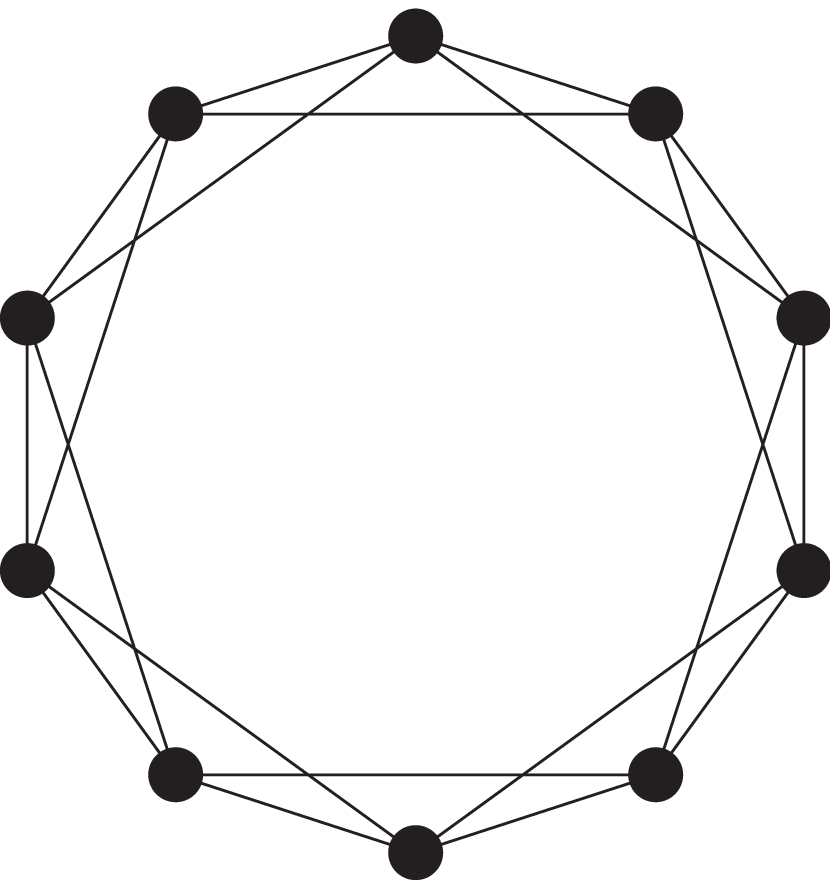}\\
    {(b) $k=2$}
   \end{center}
   \end{minipage}
  \begin{minipage}{35mm}
   \begin{center}
    \includegraphics[scale=0.3]{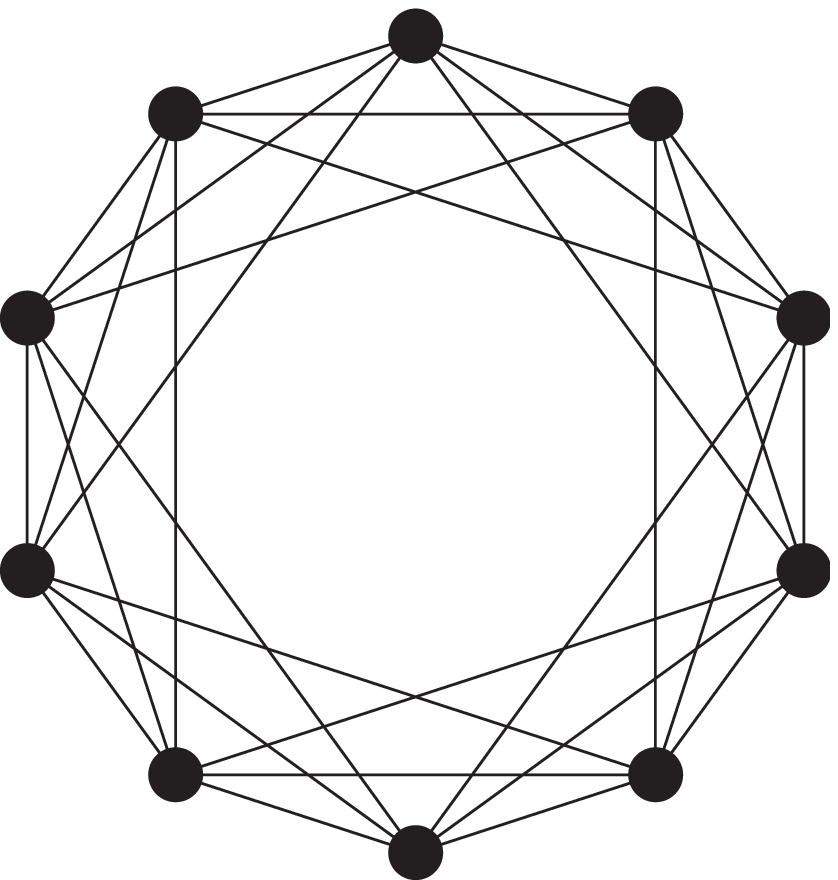}\\
    {(c) $k=3$}
   \end{center}
  \end{minipage}
  \vspace{5mm}
  \caption{Enhanced cycles with 10 vertices}
  \label{eq:enhanced_cycle}
 \end{center}
\end{figure}

\begin{figure}[!h]
 \begin{center}
  \includegraphics[scale=0.5]{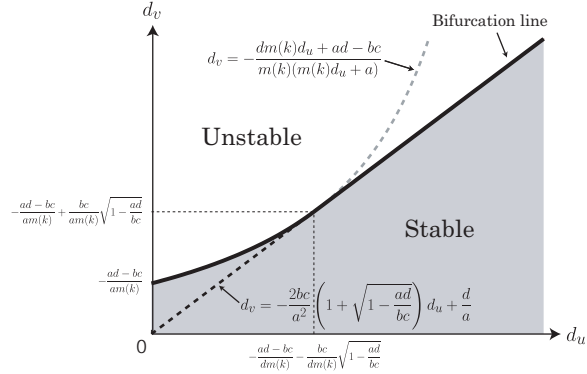}
 \end{center}
 \vspace{5mm}
 \caption{The bifurcation line of the homogeneous steady state is given by both a hyperbolic curve and a linear line when the parameters $a,b,c$, and $d\,(a>0)$ satisfy (\ref{eq:expand_condition}).}
  \label{fig:fig5.eps}
\end{figure}

From now on, we focus on an enhanced cycle with $N$ vertices as the graph $G$.
We express the enhanced cycle whose average degree is $2k\,(k\in\left\{1,2,\ldots,[(N-1)/2]\,\right\})$, as a symbol $C_{N,k}$.
To be exact, the adjacency matrix of the enhanced cycle $C_{N,k}$ is given by $A_{ij}=1\, (0<j-i+N\leq k \quad \mbox{or} \quad 0<i-j+N\leq k \mod N)$, or $=0\,(\mbox{otherwise})$.
Figure~\ref{eq:enhanced_cycle} shows three examples of the enhanced cycle $C_{N,k}$ with 10 vertices.
\\
We should note to have
\begin{equation*}
 \sigma_{C_{N,k}}(l)=\left\{\begin{array}{cl}
				 0&(l=0),\\
					-(2k+1)+\frac{\sin(2k+1)\frac{\pi l}{N}}{\sin\frac{\pi l}{N}}&(l=1,2,\ldots,N-1).
				       \end{array}\right.\label{eq:sigma_{C_k}(l)}
\end{equation*}
By using (\ref{eq:d_condition}), we can approximately analyze the stability of homogeneous steady state for a large enough number $N$ under the conditions
\begin{equation}
 \left\{\begin{array}{l}
  \max\left\{|a|, 1\right\}+d < 0,\\
	 ad-bc > 0,\\
	 (a-d)^2 + 4bc\geq 0,\\
	 a\leq 1.
	\end{array}
 \right.\label{eq:expand_condition}
\end{equation}
Assuming the parameters $a,b,c,d$ satisfies (\ref{eq:expand_condition}), we have a condition that the homogeneous steady state gets stable for the number $N$ enough larger than the number $k$ (i.e. $k\ll N$) as follows.
For $a\leq 0$, the steady state becomes stable when $d_u, d_v>0$.
On the other hand, the condition for $a>0$ is a little complicated.
To describe it, we consider a function $S_k(x)\,(x\in [0,1))$ such that
\begin{equation*}
 S_k(x)=\left\{\begin{array}{cl}
	0& (x=0),\\
	       -(2k+1)+\frac{\sin(2k+1)\pi x}{\sin\pi x}& (0<x<1),
	      \end{array}\right.
\end{equation*}
which originates from (\ref{eq:sigma_{C_k}(l)}).
Then the homogeneous steady state is stable under the condition
\begin{equation*}
 \left\{\begin{array}{ll}
  0<d_v<-\frac{d m(k) d_u+ad-bc}{m(k)(m(k)d_u+a)} & \left(0 < d_u < -\frac{ad-bc}{d m(k)}-\frac{bc}{d m(k)}\sqrt{1-\frac{ad}{bc}}\right),\\[2mm]
	0<d_v<-\frac{2bc}{a^2}\left(1+\sqrt{1-\frac{ad}{bc}}\right)d_u+\frac{d}{a} & (\mbox{otherwise}),
	\end{array}\right.
\end{equation*}
where $m(k)=\min_{x\in[0,1)}S_k(x)$.
So, when we set $a>0$, the bifurcation line of the homogeneous steady state consists of a continuous line which is produced by both a hyperbolic curve and a linear line, as shown in Figure~\ref{fig:fig5.eps}.

While we have treated the enhanced cycle in the case of $k\ll N$, we can also approximately compute the stability of the homogeneous steady state for the complete graph $K_N$ with $N$ vertices, supposing that it does not have any self-loop.
If the number $N$ is an odd integer and $k=(N-1)/2$, the enhanced cycle $C_{N,k}$ is equivalent to the complete graph $K_N$.
In the following result, we do not have to consider the assumption (\ref{eq:expand_condition}).
For the complete graph $K_N$, the eigenvalues $\lambda_{\pm}(l)\,(l=0,1,\ldots,N-1)$ of the matrix (\ref{eq:TM:diffusuin matrix}) are computed as
\begin{align*}
 \lambda_{\pm}(l)=&\left\{\begin{array}{ll}
		    \frac{a+d\pm\sqrt{(a-d)^2+4bc}}{2}&(l=0),\\[2mm]
		     \frac{h_{N,1}(d_u,d_v)\pm\sqrt{h_{N,2}(d_u,d_v)}}{2}&(l=1,2,\ldots,N-1),			   
			  \end{array}\right.
\end{align*}
where
\begin{align*}
 h_{N,1}(x,y)=&-N(x+y)+a+d,\\
 h_{N,2}(x,y)=&N^2(x-y)^2-2N(a-d)(x-y)+(a-d)^2+4bc.
\end{align*}
By using the similar method as the approximate analysis for the enhanced cycle, we see that the stability does not depend on the coefficients $d_u$ and $d_v$ as $N\gg 1$ and it is determined by the condition $\Re(\sqrt{(a-d)^2+4bc})< -(a+d)$, where $\Re(z)$ means the real part of the complex number $z$.

\subsection{Turing instability on networks generated by the threshold network model}
Let $X_{1},\ldots ,X_{N}$ be a sequence of independent and identically distributed random variables with a common distribution function $F$. If we consider the threshold network model with the vertex weights $X_{1},\ldots ,X_{N}$ and the threshold value $\theta \in \mathbb{R}$, then 
\begin{align*}
D_{N}(i)=\sum _{\substack{1\leq j\leq N\\j\neq i}}I_{\{X_{i}+X_{j}>\theta \}}
\end{align*}
is the degree of a vertex $i$, where $I_{\{X_{i}+X_{j}>\theta \}}=1$ if $X_{i}+X_{j}>\theta $ and $=0$ otherwise. Let 
\begin{align*}
\nu _{N}(dx)=\frac{1}{N}\sum_{1\leq i\leq N}\delta\left(x-\frac{D_{N}(i)}{N}\right)dx
\end{align*}
be the empirical distribution of the normalized degree sequence $D_{N}(1)/N,\ldots D_{N}(N)/N$ where $\delta (x)$ is the delta function. By the definition, the $m$-th moment of the empirical distribution $\nu _{N}(dx)$ is $(1/N)\sum_{1\leq i\leq N}(D_{N}(i)/N)^{m}$. Using the same arguments of the proof of Theorems 2 and 3 in \cite{ikm10MCAP} and the induction, we can easily show that 
\begin{align*}
\mathbb{P}\left(\lim _{n\to \infty }\frac{1}{n}\sum _{1\leq i\leq n}\left[\frac{D_{n}(i)}{n}\right]^{m}=\mathbb{E}\left[\left\{1-F(X_{1})\right\}^{m}\right]\right)=1.
\end{align*}
This implies that the empirical distribution $\nu _{n}(dx)$ converges weakly to the distribution of the random variable $1-F(X_{1})$ with probability one. 

On the other hand, the eigenvalues of the Laplacian matrix of the threshold network model are expressed as follows (see Theorems 2 and 3 in \cite{Merris1994}):
\begin{align*}
\lambda _{N}(N-i)=\sharp \left\{j : D_{N}(j)\geq N-i\right\},
\end{align*}
for $1\leq j\leq N$. Combining these two observations, we obtain the empirical distribution 
\begin{align*}
\mu _{N}(d\lambda )=\frac{1}{N}\sum_{1\leq i\leq N}\delta\left(\lambda -\frac{\lambda _{N}(N-i)}{N}\right)d\lambda
\end{align*}
converges weakly to the distribution of the random variable $1-F(X_{1})$ with probability one. 
When $X_1$ follows the exponential distribution with a parameter $\lambda$, that is, the density function 
$$
f(x)=
\begin{cases}
\lambda \mathrm{e}^{-\lambda x} & (x\ge 0),\\
0 & (x<0),
\end{cases}
$$
we have the density function on eigenvalues of the Laplacian matrix of a graph generated by the threshold network model
$$
1-F(\theta-X_1)=
\begin{cases}
\delta_1(dk) & (\theta\le 0),\\
I_{(\mathrm{e}^{-\lambda \theta},1)(k)}\cdot\frac{\mathrm{e}^{-\lambda\theta}}{k^2}+\mathrm{e}^{-\lambda\theta}\cdot \delta_1(dk) & (\theta>0). 
\end{cases}
$$
Roughly speaking, this result implies that the number of small eigenvalues is quite small, compared to the number of large ones when we consider the case $\theta>0$ which corresponds to the numerical results in section \ref{Networks generated by Threshold model}. Therefore, we can expect that the ratio of the number of small eigenvalues to the total number of eigenvalues decreases as $N$ tends to infinity. 
This insight explains the reason why the probability of destabilization of the homogeneous steady state becomes low in our numerical results in Figures \ref{SUR_TV} and \ref{figure10} as $N$ increases.

%%%%%%%%%%%%%%%%%%%%%%%%%%%%%%%%%%%%%%%

\section{Concluding remarks}
\label{Concluding remarks}
In the present paper, we dealt with the Turing instability in reaction-diffusion models defined on complex networks, and revealed that the stable-unstable regions of a homogeneous steady state differed, depending on structural properties of networks which were generated by models. We found specific properties on the Turing instability on complex networks, namely the existence of a parameter region of network connectivity induced instability in the case of the Erd\H{o}s-R\'enyi model, a variety of the stable-unstable regions under the equal average degree, and the decrease in the probability of destabilization of the homogeneous steady state according to an increase of $N$ in the threshold network model. 
In addition, we obtained some theoretical results, that is, the existence of the unstable region in the case of an enhanced ring and the result on the eigenvalue distribution for the threshold network model. In particular, as shown in Figure \ref{fig:fig5.eps} the boundary of the stable and unstable regions consists of a hyperbolic curve and a linear line. This can be explained from the numerical results on eigenvalues for $p=0$ in Figure \ref{figure8} as follows: 
The smallest eigenvalue $\lambda_1$ is equal to zero, and all the eigenvalues are distributed in the range $[0,c]$ for a positive constant $c$. Moreover, it appears that the largest eigenvalue $\lambda_N$ is bounded uniformly in $N$. These and the numerical results for $p=0$ in Figure \ref{figure8} suggest that the infinite number of eigenvalues are continuously distributed in the range $[0,\lambda_\infty]$ when $N$ tends to infinity, where $0<\lambda_\infty<\infty$. Therefore, the bifurcation curve corresponding to the eigenvalue $\lambda_\infty$ yields the hyperbolic curve and the bifurcation curves corresponding to the other eigenvalues generates the linear line in Figure \ref{fig:fig5.eps}. 
Since the Turing instability is generally the onset of a self-organized pattern formation, we can expect that an inhomogeneous pattern emerges in a self-organized way in a reaction-diffusion model defined on a network. 
\begin{figure}[htbp]
\begin{center}
\includegraphics[width=110mm]{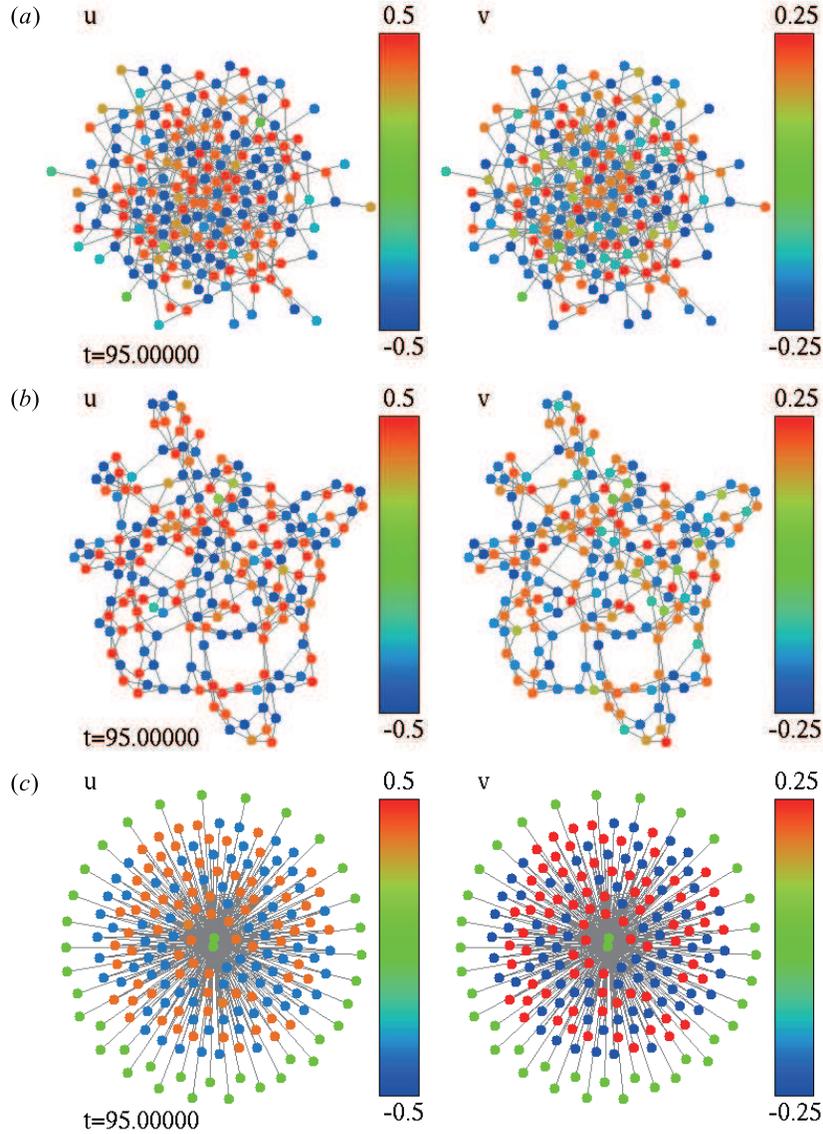}
\caption{Network-organized Turing patterns in \eqref{FHN}. The parameter values are $d_u=0.1$, $d_v=10$, $a=-1$, $\varepsilon=20$, and $\gamma=0.5$. The number of vertices $N$ is $200$. 
(a) Turing pattern on a network generated by the Erd\H{o}s-R\'enyi model with $p=0.02$. The average degree of the network is $4.22$. 
(b) Turing pattern on a network generated by the Watts-Strogatz model with $p=0.1$ and $B=4$. The average degree of the network is $4$. 
(c) Turing pattern on a network generated by the threshold network model with $\theta=6.5$. The average degree of the network is $3.86$. In each figure, the hubs are placed at the center and the vertices with small degree at the periphery. }
\label{figure18}
\end{center}
\end{figure}
Figure \ref{figure18} shows examples of a self-organized pattern on a complex network generated by the Erd\H{o}s-R\'enyi, the Watts-Strogatz, and the threshold network models, respectively. Here, we numerically solve the following model defined on a network with the FitzHugh-Nagumo type nonlinearity: 
\begin{equation}
\label{FHN}
\begin{aligned}
\frac{du_i}{dt}&= -d_u\sum_{j=1}^{N} L_{ij} u_j+u_i(1-u_i)(u_i-a)-v_i,\\
\frac{dv_i}{dt}&= -d_v\sum_{j=1}^{N} L_{ij} v_j +\varepsilon(u_i-\gamma v_i), 
\end{aligned}
\qquad (i=1,2,\cdots, N),
\end{equation}
where $a$, $\varepsilon$, and $\gamma$ are constants, and the coefficients $d_u$, $d_v$, and the Laplacian matrix $L_{ij}$ are defined in section \ref{Formulation of Reaction-Diffusion model on networks}. In the figure, the vertices are arranged on the periphery of a circle and their colors correspond to values shown in the indicator bar, and the lines between two vertices mean links. The left and the right figures respectively denote the quantities for $u$ and $v$. 
Figures \ref{figure18} (a), (b), and (c) represent stationary states of \eqref{FHN} on a network generated by the Erd\H{o}s-R\'enyi, the Watts-Strogatz, and the threshold network models, respectively. We can observe network organized patterns as a consequence of the Turing instability on a network in any cases. 
However, we do not have any more results on the network-organized patterns. 
A further investigation between Turing instability discussed in the present paper and network-organized patterns is our future problem. 
The existence of such network-organized patterns was numerically discussed in \cite{NM} and the authors refered to several features of network-organized pattern formation. Moreover, recently, studies on a stable inhomogeneous pattern with a single differentiated node have proceeded from the viewpoint of bifurcation analysis in \cite{W}. As another typical solution in reaction-diffusion equations on continuous media, a traveling wave solution is well known, which moves with constant speed without changing the profile. In \cite{KKM, IHHS, IS, KIMS}, wave-like phenomena were also observed in reaction-diffusion models on networks. Interestingly, the occurrence of propagation failure of waves and pinned waves in network organized reaction-diffusion models were reported. However, a lot of things on the network-reaction-diffusion models are not clear. It is not easy to understand such phenomena arising in a reaction-diffusion model on networks due to stochastic construction of a network and the large number of equations.

Though we dealt with three types of models, i.e. the Erd\H{o}s-R\'enyi, the Watts-Strogatz, and the threshold network models, many models have been proposed besides the three models, for instance the Bar\'abasi-Albert model. Numerical and theoretical investigation of a relation between such models and the stable-unstable regions of the homogeneous steady state is a future work. A challenging problem is to treat a nonlinear system with the Laplacian matrix of a graph such as \eqref{FHN}. In this paper, since we focused on the Turing instability of reaction-diffusion models on networks, the associated linear systems with network architecture were considered. However, in order to understand the mechanism of network-organized pattern formation, we need to analyze the nonlinear system instead of the linear system. 
Moreover, from the viewpoint of mathematical ecology, a two patches model was proposed in \cite{AF2}, where the migration rate of each species was influenced by its own and the other one's density, that is, the model on two patches possesses the cross-diffusion-like effect which is one of the nonlinear diffusion effects. The extension of the model on two patches which locate on complex networks may contribute to the understanding of the spatially distribution of biological species on patchy environments. Applications of reaction-diffusion models on complex networks for specific problems are also a future work.

\section*{Acknowledgement}
\noindent 
HI is grateful to JSPS KAKENHI 26800084 for the support. 
TM is grateful to the Japan Society for the Promotion of Science for the support and the Math. Dept. UC Berkeley for hospitality, and thanks Y. Kitada for a useful comment.

\end{document}